\documentclass[12pt]{article}
\usepackage{epsf}
\usepackage{a4wide,graphicx}
\usepackage{cite}

\newcommand{\mev}{\textrm{ MeV}}
\newcommand{\be}{\begin{equation}}
\newcommand{\ee}{\end{equation}}
\newcommand{\ba}{\begin{eqnarray}}
\newcommand{\ea}{\end{eqnarray}}
\newcommand{\nn}{\nonumber}

\begin{document}

\title{Quantum loops in radiative decays of the $a_1$ and $b_1$
 axial-vector mesons}

\author{
L.~Roca$^1$, A.~Hosaka$^2$ and E.~Oset$^3$\\
{\small{\it $^1$Departamento de F\'{\i}sica. Universidad de Murcia. E-30071
Murcia.  Spain}}\\
{\small{\it $^2$Research Center for Nuclear Physics (RCNP), Ibaraki, Osaka
  567-0047, Japan}}\\
{\small{\it $^3$Departamento de F\'{\i}sica Te\'orica and IFIC,
Centro Mixto Universidad de Valencia-CSIC,}}\\
{\small{\it Institutos de
Investigaci\'on de Paterna, Aptdo. 22085, 46071 Valencia, Spain}}
}

\date{\today}

\maketitle

 \begin{abstract} 
 
A previous model where the low-lying axial-vector mesons are
dynamically generated, implementing unitarity in
coupled channels in the vector-pseudoscalar ($VP$) meson interaction,
is applied to evaluate the decay widths of the
$a_1(1260)$ and $b_1(1235)$ axial-vector mesons into $\pi\gamma$.
Unlike the case of the $a_1$, the $b_1$ radiative decay is systematically
underestimated at tree level.  In this work we evaluate for the first 
time the loop
contribution coming from an initial $VP$ vertex. Despite the 
large superficial divergence of the loops, the convergence of the relevant 
loops can be established by using arguments of gauge invariance.
 The partial
decay widths obtained agree with the experimental values
within uncertainties, and we show that the loop contribution is
crucial in the $b_1$ case and also important for the $a_1$ case.

\end{abstract}

\section{Introduction}  The unitary extensions of chiral
perturbation theory ($\chi PT$) have allowed to extend the range of
energies where the hadron interaction can be studied.  At the same
time they have also shown that many meson and baryon resonances are
dynamically generated and can be interpreted as quasibound states
of pairs of hadrons in coupled channels \cite{review}. A case very
well studied is the one of the interaction of the octet of
pseudoscalar mesons  \cite{npa,oop,kaiser,markushin,nieves} from
where a nonet of scalar mesons are generated.  Much less studied is
the case of the interaction of vector mesons  with pseudoscalar
mesons, where two independent works \cite{kolo,Roca:2005nm} have 
shown that the axial vector mesons can be generated dynamically. 
This novel idea should be confronted with experiment to test the
accuracy of its predictions.  Some of these predictions have
already been tested in Ref.~\cite{Geng:2006yb}. Contrary to other
pictures like quark models, where external sources are coupled to
the quarks, in the dynamically generated picture one assumes that
the largest weight of the wave function is due to the two meson
cloud, and consequently, the coupling of external sources proceeds
via the coupling to the meson components.  One interesting test
which brings light into this issue is the radiative decay of the
resonances. This is the purpose of the present work where we
concentrate on the radiative decay of the $b_1^+$ and $a_1^+$ axial
vectors into $\pi^+ \gamma$. The $a_1^+$ radiative decay has been
studied within different contexts, for instance vector meson
dominance is used in \cite{shuryak,haglin}, relating the radiative
decay with the $\rho \pi$ decay of the $a_1^+$.  Chiral Lagrangians
with vector meson dominance (VMD) are also used in
\cite{Ecker:1988te} to obtain the radiative width of $a_1^+ \to
\pi^+\gamma$.  An SU(3) symmetric Lagrangian is used in
\cite{palomar} to account for strong decays of the axial vector
mesons and by means of  VMD the amplitude for $a_1^+ \to
\pi^+\gamma$ is studied and related to the one of
\cite{Ecker:1988te}. A common feature of these works is that the 
$b_1^+ \to \pi^+\gamma$  reaction is not discussed and its
evaluation in  \cite{palomar} using VMD along the same lines as
the  $a_1^+ \to \pi^+\gamma$ gives rise to a decay rate
substantially smaller than experiment. The  $b_1^+ \to \pi^+\gamma$
decay is also neglected in the analysis of \cite{haglin} citing 
the small rates obtained.

  The rates of $a_1^+ \to \pi^+\gamma$ and $b_1^+ \to \pi^+\gamma$ are also
evaluated in  \cite{rosner} using quark models for the $a_1 \to \pi \rho$ and 
$b_1 \to \pi \omega$ and VMD to relate these amplitudes with the radiative
decay.  It is emphasized there that because of the factor 1/3 of the $\omega
\gamma$ coupling relative to the one of $\rho \gamma$ there is a reduction
factor
of 1/9 for the radiative decay $b_1^+ \to \pi^+\gamma$ compared to that of the 
$a_1^+ \to \pi^+\gamma$ decay, resulting in a ratio of these two rates in
contradiction with experiment (this is the same argument found in
\cite{palomar} as responsible for the small rate of the $b_1^+ \to \pi^+\gamma$
decay).  

  In the present work we shall also use the tree level VMD
 amplitudes, but in addition, the nature of the axial vector mesons
 as dynamically generated resonance provides a strong coupling to
 $K^* \bar{K}$ and $\bar{K^*} K$, and subsequent loops with these
 intermediate states and the photon emitted from these constituents
 should be considered.  We show that the loops are very important,
 particularly for the case of  the $b_1^+ \to \pi^+\gamma$ decay,
 and the simultaneous consideration of the VMD amplitudes at tree
 level and loop
 contributions leads to a good description of both radiative
 decays.  

     We shall also show some technical details involving loops with
vector mesons. Using arguments of gauge invariance and the Feynman
parametrization, one can prove that the loops involving one vector
meson and two pseudoscalar mesons are finite, in spite of the large
degree of superficial divergence. This was found in
\cite{lucio,close,Marco:1999df,oller} with the loops involved in
radiative decay of the $\phi$ containing three pseudoscalar mesons.

\section{Formalism}

In Ref.~\cite{Roca:2005nm} most of  the low-lying axial vector
mesons were dynamically generated from the s-wave interaction of
the octet of vector-mesons with the octet of pseudoscalar-mesons by
using the techniques of the chiral unitary theory.  With the only
input of a chiral Lagrangian for a vector and pseudoscalar (VP)
mesons and the implementation of unitarity
in coupled channels, these resonances show up as poles in the
second Riemann sheet of the unitarized scattering amplitudes. By
evaluating the residues of the scattering amplitudes at the pole
positions, the couplings of the dynamically generated axial-vector
resonances to the different $VP$ channels can be obtained. By using
these couplings, we found a nice agreement with the experimental $VP$
partial decay rates, despite the fact that no parameters were
fitted to experimental data of the axial-vector mesons.

In view of the dominant contribution of the $VP$ channels in the
building up and decay of the axial-vector resonances,  our starting
point to study the radiative decay of the $b_1$ and $a_1$ is the
transition of these resonances into the allowed $VP$ channels and 
attaching the photon to the relevant meson lines and vertices. 
\begin{figure}[h]
 \begin{center}
\includegraphics[width=0.8\textwidth]{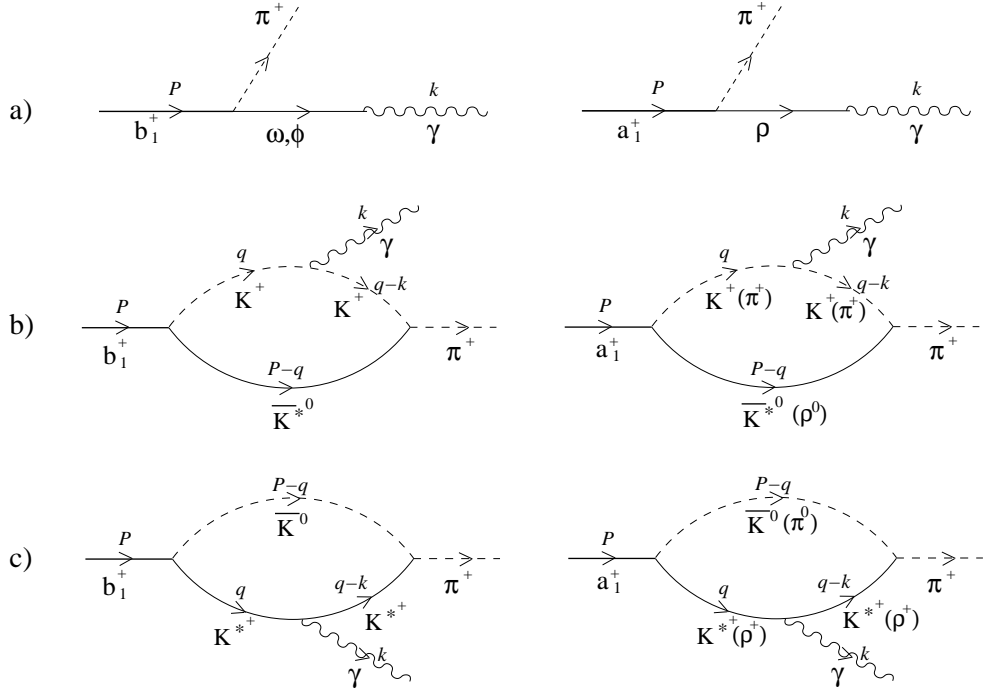}
\caption{Feynman diagrams contributing to the radiative
axial-vector meson decay.} 
\label{fig:diagrams}
 \end{center}
\end{figure} 
The first kind of mechanisms considered are the tree
level vector meson dominance (VMD) contributions, shown in
Fig.~\ref{fig:diagrams}$a)$. Furthermore, the radiative decay can
also proceed from loops of the $VP$ pair with the photon emitted
from either the pseudoscalar or the vector meson leg,
Fig.~\ref{fig:diagrams}$b)$ and $c)$. A diagram with the photon
directly emitted from the $VPP$ vertex is 
needed to ensure gauge invariance, but we
will explain later on
(after Eq.~(\ref{eq:finaltloops}))
 that, using arguments of gauge invariance, we
do not  need to evaluate it directly.  On the other hand, another
kind of loops  containing the $VP\gamma$ and $VVP$ vertices is also
possible, however, they involve two abnormal intrinsic parity
vertices, and hence its contribution should be rather small
compared to those already considered.  This is indeed the case in
the analogous loops present in the radiative  decay of the $\phi$ 
meson, as it was found  in \cite{Palomar:2003rb}.

The intermediate $VP$ states possible in the loops are those used
in Ref.~\cite{Roca:2005nm} to build up the axial vector mesons.
These are, for the $b_1$: $1/\sqrt{2}(\bar{K^*} K+K^* \bar{K})$,
$\phi\pi$, $\omega\pi$, $\rho\eta$, and for the
 $a_1$:$1/\sqrt{2}(\bar{K^*} K - K^* \bar{K})$, $\rho\pi$. Note
 however, that the coupling of the $\phi\pi$, $\omega\pi$,
 $\rho\eta$ to the final pion violates $G-$parity and hence these
 channels do not contribute to the $b_1$ radiative decay. Thus,
 only the diagrams in Fig.~\ref{fig:diagrams} must be evaluated.

Let us start with the evaluation of the tree level contributions.
For the $V\gamma$ vertex we use the amplitude
\be
t_{V\gamma}=-e \lambda_V\, F_V\, M_V\, \epsilon_V\cdot\epsilon
\ee
with $\lambda_V=1$, $1/3$, $-\sqrt{2}/3$ for $\rho$, $\omega$ and
$\phi$ respectively, $F_V=156\pm 5$MeV \cite{Palomar:2003rb},
 $M_V$ is the vector
meson mass and $\epsilon_V$ and $\epsilon$ are the vector-meson and
photon polarization vectors respectively, and $e$ is taken 
positive.

The axial-vector meson coupling to $VP$ can be expressed
\cite{Roca:2005nm} as 
\be
t_{AVP}=g_{AVP} \, \epsilon_A\cdot\epsilon_V
\ee 
\noindent
where $\epsilon_A$ is the axial-vector meson polarization vector.
The couplings $g_{AVP}$ are obtained in Ref.~\cite{Roca:2005nm}
by evaluating the residues at
the poles of the $VP$ unitarized scattering amplitudes
 and are given in table~VII of that reference.
 Note that in Ref.~\cite{Roca:2005nm} the couplings are
 given in isospin
base and for given $G-$parity states, hence the 
appropriate projection to charge base has to be done.
 In Ref.~\cite{Roca:2005nm} no theoretical errors were quoted for
these $AVP$ couplings. However, for the purpose of evaluating the
theoretical uncertainty in the calculations of the present work, we
have estimated the uncertainties in these couplings in the
following way: for the $b_1$ case, we have considered the change in
the couplings due to a resonable uncertainty of $10$\% in the
 only free 
parameter of the
model, the subtraction constant $a\sim-1.85$ (see Ref.~\cite{Roca:2005nm}
for details).  We consider further
uncertainties from changing $f$, as will be explained after 
Eq.~(\ref{eq:LVPP}). 
For the $a_1$ case, in
Ref.~\cite{Roca:2005nm} the mass obtained was
$1011\mev$, somewhat below the nominal mass in the PDG
\cite{pdg06}, $1230\mev$. (Note however that the total width is
$250-600\mev$ in the PDG, which gives an idea of the uncertainty in
the mass). In Ref.~\cite{Roca:2005nm}, the mass was obtained with a
value of $a=-1.85$.
If we use the value $a=-1.1$ and $f=f_K$,
 we obtain a mass closer
to the nominal of $1080\mev$, and it is not easy to get larger
mass. In this case the coupling to $\rho \pi$, the dominant
channel, is increased by $\sim 25$\%. From there, we get an idea of
the uncertainties in the $a_1$ couplings.

This leads to the following radiative decay amplitudes for the
tree level diagrams, (Fig.~\ref{fig:diagrams}$a)$):

\ba
t_{b_1^+\to
\pi^+\gamma}&=&\frac{1}{3}e F_V
\left(\frac{1}{M_\omega}g_{b_1\omega\pi}
-\frac{\sqrt{2}}{M_\phi}g_{b_1\phi\pi}\right) \epsilon'\cdot\epsilon\nonumber\\
t_{a_1^+\to
\pi^+\gamma}&=&-\frac{1}{\sqrt{2}}e F_V
\frac{1}{M_\rho}g_{a_1\rho\pi}\epsilon'\cdot\epsilon
\ea

In the evaluation of the loops  an apparent problem arises
given the large superficial divergence due to  the loop momentum
dependence of the vertices and the $q^\mu q^\nu/M_V^2$ terms of the
vector meson propagators. However, we will explain in detail how one can
circumvent this problem invoking gauge invariance and
using a suitable Feynman parametrization of the loop integrals.

Since the only external momenta available are 
 $P$ (the axial-vector meson momentum)  and $k$
  (the photon momentum), 
 the general expression of the amplitude can be written as 
 
 \be
 T={\epsilon_A}_\mu \epsilon_\nu T^{\mu\nu}
 \label{eq:Tgeneral}
 \ee 
with
\be
T^{\mu\nu}=a\, g^{\mu\nu} + b\, P^\mu P^\nu + c\, P^\mu k^\nu
 +d\, k^\mu P^\nu + e\, k^\mu k^\nu
 \label{eq:Tmunuterms}
\ee
Note that, due to the Lorentz condition, ${\epsilon_A}_\mu P^\mu=0$,
${\epsilon}_\nu k^\nu=0$, all the terms in Eq.~(\ref{eq:Tgeneral})
vanish except for the $a$ and $d$ terms. On the other hand, gauge
invariance implies that $T^{\mu\nu}k_\nu=0$, from where one gets
\be
a=-d\,P\cdot k.
\label{eq:ad}
\ee
This is obviously valid in any reference frame, however, in the axial-vector
meson rest frame and taking the Coulomb gauge for the photon, only
the $a$ term survives in  Eq.~(\ref{eq:Tgeneral}) since 
$\vec P=0$ and $\epsilon_0=0$.
This means that, in the end, we will only need the $a$ coefficient
for the evaluation of the process. However, the $a$ coefficient can
be evaluated from the $d$ term thanks to Eq.~(\ref{eq:ad}). The
advantage of doing this is that 
there are few mechanisms contributing to the $d$ term and by dimensional 
reasons the number of powers of the loop momentum  in the numerator
 will be  reduced, as will be
clearly manifest from the discussion below. 

Let us start by evaluating the diagram ~\ref{fig:diagrams}$b)$
 with the photon
emitted from the pseudoscalar leg for the
$b_1^+\to\pi^+\gamma$ decay channel (the other channels are totally
analogous). We will call this diagram type-b, in contrast with the
type-c, with the photon attached to the vector-meson leg which will
be evaluated later on, (see Fig.~\ref{fig:diagrams}).

For the evaluation of this diagram we also 
need the $VPP$ and $PP\gamma$
vertices. The  $VPP$ Lagrangian used is (see Ref.~\cite{Palomar:2002hk} for
normalizations used)

\be
{\cal L}_{VPP}=-i\frac{g}{\sqrt{2}}\langle V^\mu 
[\partial_\mu P,P]\rangle,
\label{eq:LVPP}
\ee
where $V_\mu$ and $P$ are the usual $SU(3)$ matrices containing the
vector and pseudoscalar mesons.
In Eq.~(\ref{eq:LVPP}) $\langle ... \rangle$ 
means $SU(3)$ trace and $g=-M_V G_V/f^2$, where $M_V$ is the vector
meson mass, $G_V=55\pm 5$~MeV \cite{Palomar:2003rb} and $f$ is the
pion decay constant that we take from $93$~MeV to $1.15\times 93$~MeV
 to
take into account the uncertainty due to the use of $f$ instead of 
 $f_K$ which could actually enter some of the expressions. These 
 uncertainties in the parameters, together with
the uncertainties in other couplings of the theory, will be taken
into account later on in the evaluation of the theoretical
uncertainties of our results.

The $PP\gamma$ vertex is easily obtained from the lowest order
$ChPT$ Lagrangian \cite{Gasser:1984gg}
\be
{\cal L}=\frac{f^2}{4}\langle D_\mu U^\dagger D^\mu U\rangle
\ee
where the photon field appears in the covariant derivative.  

The amplitude for the type-b mechanism 
(see Fig.~\ref{fig:diagrams}$b$) for the
 $b_1^+\to\pi^+\gamma$ reads:

\ba
-it_b&=&-i\frac{1}{\sqrt{2}}g_{b_1K^*K} 
\int\frac{d^4q}{(2\pi)^4}\,\epsilon_A^\mu
\frac{i}{q^2-m_K^2}\frac{i}{(q-k)^2-m_K^2}\nn\\
&\times&\frac{i}{(P-q)^2-{m^2_{K^*}}} 
\left(-g_{\mu\alpha}+\frac{(P-q)_\mu(P-q)_\alpha}{{m^2_{K^*}}}\right)
\nn\\
&\times&i\frac{m_{K^*} G_V}{\sqrt{2}f^2}(k-q-p_\pi)^\alpha
(-ie)\epsilon^\nu(q+q-k)_\nu\nn\\
&=&-\frac{1}{\sqrt{2}}g_{b_1K^*K} e\frac{m_{K^*} G_V}{\sqrt{2}f^2}
\, \epsilon_A^\mu\epsilon^\nu\,I_{\mu\nu}
\label{eq:itb}
\ea
with
\ba
I_{\mu\nu}&=&\int\frac{d^4q}{(2\pi)^4}
\frac{1}{q^2-m_K^2}\frac{1}{(q-k)^2-m_K^2}
\frac{1}{(P-q)^2-{m^2_{K^*}}} \nn\\
&\times&
\left(-(2k-q)_\mu-q_\mu\frac{\left[(P-q)\cdot(2k-2P)+(P-q)^2\right]}
{{m^2_{K^*}}}\right)2q_\nu.
\label{eq:Imunu}
\ea
In Eq.~(\ref{eq:itb}), $g_{b_1K^*K}$ is the coupling of the $b_1$ to
the $K^*\bar K$ and $\bar K^* K$ G-parity positive combination as
defined in Ref.~\cite{Roca:2005nm}.
By looking at Eq.~(\ref{eq:Imunu}) one can see that in the worst
case the loop integral, as it is written, is quadratically divergent.
 At this
point, we can take advantage of the fact
that we only need to evaluate the contribution to the 
$d$ term of Eq.~(\ref{eq:Tmunuterms}), as
explained above. In fact, the most divergent term, the one 
with $(P-q)^2$, does not
contribute to the that term. Indeed, we can write
\be
(P-q)^2=(P-q)^2-{m^2_{K^*}}+{m^2_{K^*}},
\label{eq:kk1}
\ee
and then the two first addends of the right hand side of the above
equation give
\be
\int\frac{d^4q}{(2\pi)^4} \
\frac{(P-q)^2-{m^2_{K^*}}}{(P-q)^2-{m^2_{K^*}}} \ 
\frac{1}{q^2-m_K^2}\frac{1}{(q-k)^2-m_K^2} \ q_\mu q_\nu,
\ee
which does not depend explicitly on $P$,
since the $(P-q)^2-{m^2_{K^*}}$ cancels the propagator
where the $P$ appears and, therefore, this integral cannot give contribution
to the $d$ term. 
Hence, for the purpose of evaluating the $k_\mu P_\nu$ contribution,
Eq.~(\ref{eq:Imunu}) can be written as
\ba
I_{\mu\nu}&=&\int\frac{d^4q}{(2\pi)^4}
\frac{1}{q^2-m_K^2}\frac{1}{(q-k)^2-m_K^2}
\frac{1}{(P-q)^2-{m^2_{K^*}}} \nn\\
&\times&\left(-4k_\mu+2q_\mu \left[1-
\frac{(P-q)\cdot(2k-2P)+{m^2_{K^*}}}
{{m^2_{K^*}}}\right]\right)q_\nu
\label{eq:Imunub}
\ea
which has one power less in the variable $q$ 
than  Eq.~(\ref{eq:Imunu}).
Next we use the Feynman parametrization to evaluate this integral and
we will see that the contribution to the $d$ term is convergent.

We use the identity
\be
\frac{1}{abc}=2\int_0^1 dx\int_0^x dy 
\frac{1}{[a+(b-a)x+(c-b)y]^3}.
\ee 
By setting
\ba
a&=&(P-q)^2-{m^2_{K^*}},\nn\\
b&=&q^2-m_K^2,\nn\\
c&=&(q-k)^2-m_K^2
\ea
and performing the change of variable
\be
q=q'+[(1-x)P+yk],
\ee
we have
\ba
I_{\mu\nu}&=&2\int_0^1 dx\int_0^x dy\int\frac{d^4q'}{(2\pi)^4}
\frac{1}{({q'}^2+s+i\varepsilon)^3}
\Bigg[-4k_\mu(q'_\nu+[(1-x)P_\nu+yk_\nu])\nn\\
&+&2(q'_\mu + [(1-x)P_\mu+yk_\mu])
(q'_\nu +[(1-x)P_\nu+yk_\nu])\nn\\
&\times&\left(
1-\frac{(P-q'-[(1-x)P+yk])\cdot(2k-2P)+{m^2_{K^*}}}
{{m^2_{K^*}}}\right)\Bigg],
\label{eq:kk2}
\ea
with
\be
s=(1-x)(xM_{b_1}^2-{m^2_{K^*}}-2y P\cdot k)-x m_K^2.
\label{eq:s}
\ee

However, in Eq.~(\ref{eq:kk2}), all the terms that contribute to the 
$d$ coefficient are finite. Hence, in the end, the evaluation
of the amplitude of the type-b diagram is
 completely finite, and gives:

\ba
t_b&=&-\frac{1}{\sqrt{2}}g_{b_1K^*K} e\frac{m_{K^*} G_V}{\sqrt{2}f^2}
2 P\cdot k \, \epsilon_A\cdot\epsilon \nn\\
&\times&\int_0^1 dx\int_0^x dy \frac{1}{32\pi^2}\frac{1}{s+i\varepsilon}
\left[-4(1-x)-4 y(1-x)\frac{(xP-yk)\cdot(k-P)}{{m^2_{K^*}}}\right]
\label{eq:finalta}
\ea
In the derivation of Eq.~(\ref{eq:finalta}) from 
Eq.~(\ref{eq:kk2}) we have used the relation between the $a$ and $d$
coefficients given by Eq.~(\ref{eq:ad}). We have also used that
\cite{mandl}
\be
\int d^4q'\frac{1}{(q'^2+s+i\varepsilon)^3}
=i \frac{\pi^2}{2}\frac{1}{s+i\varepsilon} 
\ee
and that
terms with odd powers of $q'$ vanish when performing the
integration, by symmetry reasons. It is also worth explaining a 
subtle cancellation which occurs 
 between two logarithmic divergent pieces when deriving 
  Eq.~(\ref{eq:finalta}), as 
explained below:

In Eq.~(\ref{eq:kk2}), apart from the terms which provide a finite
contribution to the $d$-term, already considered in Eq.~(\ref{eq:finalta}),
 there are two more terms which contribute to the $d$-term and which 
 are logarithmically divergent. One of them  goes as $yk_\mu
q'_\nu q'_\alpha (k-P)^\alpha$. After the $q'$ integration, this
gives a term proportional to $-y k_\mu P_\nu$, since the 
$q'_\nu q'_\alpha$ gives a result proportional to 
$g_{\nu\alpha}$. The other term goes as
$q'_\mu(1-x)P_\nu q'_\alpha(k-P)^\alpha$ and gives a term 
proportional to $(1-x)k_\mu P_\nu$ after the $q'$ integration, 
 with the same proportionality
coefficient as in the other case. However after doing the $x$ and $y$
integration these two terms give the same result but 
with opposite sign.
 Hence these two possible 
sources of divergent contribution to the $k_\mu P_\nu$ cancel
exactly among themselves.
 Therefore the expression of the amplitude in 
Eq.~(\ref{eq:finalta}) is totally finite. 
It is worth stressing again the power  of the technique used
 here to
evaluate the amplitude coming from the type-b loops
since, despite starting from a loop
quadratically divergent, we have been able to get rid of all the
divergences in an exact way.

At this point, it is worth noting that the numerical evaluation
of the term proportional to $1/{m^2_{K^*}}$ in 
Eq.~(\ref{eq:finalta}) is about $5$\% of the other term. This term
comes from the last factor of Eq.~(\ref{eq:kk2})
which essentially is due to the
the $p_V^\mu p_V^\nu/{m_V}^2$ part of the vector meson
propagator.
 Hence, the $1/{m_V}^2$
terms can be safely ignored in the evaluation of the type-c diagram.
This approximation is expected to be very accurate since, advancing
some results, the type-c diagrams will be found to be very small
compared to the type-b and hence the $1/{m_V}^2$ is a small piece of
a diagram contributing little to the radiative decay width.
Nonetheless, we will
include later on this uncertainty in the theoretical error analysis. 

Now we evaluate the amplitude corresponding to the type-c diagram,
Fig.~\ref{fig:diagrams}$c)$.

We also need in these case the $VV\gamma$ vertex that we get from
 gauging the charged vector meson kinetic term

\be
{\cal L}=-\frac{1}{2}F^\dagger_{\mu\nu}F^{\mu\nu}
\ee
with $F^{\mu\nu}=\partial^{\mu}  V^{\nu} 
- \partial^{\nu}  V^\mu$ with the minimal coupling
substitution $\partial_\mu\to \partial_\mu +i q A_\mu$.

After neglecting the  $1/{m_V}^2$ 
term of the vector meson propagator
by the reasons explained above, we have:

 \ba
-it_c=\frac{1}{\sqrt{2}}g_{b_1K^*K} e\frac{m_{K^*} G_V}{\sqrt{2}f^2}
\, \epsilon_A^\mu\epsilon^\nu\,J_{\mu\nu}
\label{eq:kk3}
\ea
 with
\ba
J_{\mu\nu}&=&\int\frac{d^4q}{(2\pi)^4}
\frac{1}{q^2-{m^2_{K^*}}}\frac{1}{(q-k)^2-{m^2_{K^*}}}
\frac{1}{(P-q)^2-{m_K}^2} \nn\\
&\times&\left[2q_\nu (2P-k-q)_\mu-q\cdot(2P-k-q)g_{\mu\nu}
-(q-k)_\mu (2P-k-q)_\nu\right].
\label{eq:Jmunu}
\ea
After keeping only the terms contributing to $k_\mu P_\nu$,
doing the Feynman parametrization and using the relation of
Eq.~(\ref{eq:ad})
the final expression of the amplitude coming from the type-c diagram
is

\ba
t_c=-\frac{1}{\sqrt{2}}g_{b_1K^*K} e\frac{m_{K^*} G_V}{\sqrt{2}f^2}
2 P\cdot k \, \epsilon_A\cdot\epsilon 
\int_0^1 dx\int_0^x dy \frac{1}{32\pi^2}\frac{1}{s'+i\varepsilon}
\left[y(1-x)-3x+2y+1\right],
\label{eq:finaltb}
\ea
with
\be
s'=(1-x)(x M_{b_1}^2-{m_K}^2-2y P\cdot k)-x {m^2_{K^*}}.
\label{eq:sp}
\ee

 Despite the  smallness of the terms coming from the 
$1/{m_V}^2$ term of the vector meson propagator in the type-b 
mechanism, it is worth
mentioning a technicality regarding the cancellations of the
divergences had we evaluated these terms in the type-c loop.
 If one keeps 
these $1/{m_V}^2$ terms in the vector-meson propagators 
one 
obtains that the terms with $1/{m_V}^4$ do not contribute to the 
$d\, k^\mu P^\nu$ term and there remains a 
logarithmic divergence proportional
 to $1/{m_V}^2$. 
 This divergence should be
expected to cancel had one included suitable tadpoles which could
cancel the offshellness of the momentums of the vector-meson in the
loops, in a similar way to what was shown in \cite{Cabrera:2000dx},
where the  factorization of the $\vec q\,^2$ terms in the loop was 
justified.  For the same reasons, this factorization was also used 
in \cite{Roca:2005nm}. Should one take this prescription here,
the  $1/{m_V}^2$ terms would be also very small. In any case we
will make  a conservative estimate of the errors induced by considering
these terms in one way or another. By knowing that the contribution 
to the width of the type-c loop  diagram is one order of magnitude 
smaller than that of the type-b  one, and that the changes found for 
the loop of type-b are of the order of 5\%, an estimate of 10\% error
of the radiative width coming from these considerations is a safe estimate.

Adding both type-b and type-c loops, we have:

\ba
t_{loops}&=&-\frac{1}{\sqrt{2}}g_{b_1K^*K} e\frac{m_{K^*} G_V}
{\sqrt{2}f^2}
2 P\cdot k \, \epsilon_A\cdot\epsilon 
\int_0^1 dx\int_0^x dy \frac{1}{32\pi^2}\nn\\
&\times&
\left(\frac{-4}{s+i\varepsilon}(1-x)[1+y(xP-yk)\cdot(k-P)/{m^2_{K^*}}]
\right.\nn\\
&+&\left.
\frac{1}{s'+i\varepsilon}
[y(1-x)-3x+2y+1]\right).
\label{eq:finaltloops}
\ea

Another possible diagram with the photon
directly emitted from the $VPP$ vertex, which is needed to ensure
gauge invariance of the set of diagrams, 
 does not give contribution to the $d$
coefficient since the vertices involved are both of the type
$\epsilon'\cdot\epsilon$, with $\epsilon'$ either the vector or
axial-vector meson polarization vector. Therefore there is no $k$
momentum dependence either in the vertices or in the propagators and
hence the integration cannot give contribution to the $k_\mu P_\nu$
structure.

Concerning the $a^+_1\to \pi^+\gamma$ decay, the evaluation is totally
analogous to the previous one. The amplitude from the $K^*K$ loops,
when adding both type-b and -c mechanisms is:

\ba
t_{loops}^{(K^*K)}&=&-\frac{1}{\sqrt{2}}g_{a_1K^*K} e\frac{m_{K^*} G_V}
{\sqrt{2}f^2}
2 P\cdot k \, \epsilon_A\cdot\epsilon 
\int_0^1 dx\int_0^x dy \frac{1}{32\pi^2}\nn\\
&\times&
\left(\frac{-4}{s+i\varepsilon}(1-x)
[1+y(xP-yk)\cdot(k-P)/{m^2_{K^*}}]\right.\nn\\
&-&\left.\frac{1}{s'+i\varepsilon}
[y(1-x)-3x+2y+1]\right)
\label{eq:finaltloopsa1a}
\ea
where one has to change $m_{b_1}$ by $m_{a_1}$ in
the evaluation of $s$ and $s'$ in  Eqs.~(\ref{eq:s}) and
(\ref{eq:sp}).
Note the relative minus sign in the terms with $s'$
 of  Eq.~(\ref{eq:finaltloopsa1a}) with respect to 
 Eq.~(\ref{eq:finaltloops}). This is due to the fact that, as we
 mentioned above, the $b_1$ couples to the positive $G-$parity
 combination $(\bar{K^*} K+K^* \bar{K})$ while the $a_1$
couples to the negative $G-$parity
 combination $(\bar{K^*} K-K^* \bar{K})$.

In the $a_1$ case there is also the possibility of having $\rho$ and
$\pi$ in the loops. This mechanism gives:

\ba
t_{loops}^{(\rho \pi)}&=&g_{a_1\rho\pi} e\frac{m_\rho G_V}
{\sqrt{2}f^2}
2 P\cdot k \, \epsilon_A\cdot\epsilon 
\int_0^1 dx\int_0^x dy \frac{1}{32\pi^2}\nn\\
&\times&
\left(\frac{-4}{s+i\varepsilon}(1-x)
[1+y(xP-yk)\cdot(k-P)/{m_\rho}^2]\right.\nn\\
&-&\left.\frac{1}{s'+i\varepsilon}
[y(1-x)-3x+2y+1]\right)
\label{eq:finaltloopsa1b}
\ea
where one has to change $m_{b_1}$ by $m_{a_1}$, $m_{K^*}$ by
$m_\rho$, and $m_K$ by $m_\pi$ in
the evaluation of $s$ and $s'$ in  Eqs.~(\ref{eq:s}) and
(\ref{eq:sp}).

The expression of the radiative decay width in the axial-vector
meson rest frame is

\be
\Gamma(M_A)=\frac{|\vec k|}{12\pi M_A^2}|T|^2
\ee
where $M_A$ stands for the mass of the decaying axial-vector meson
and $T$ is the sum of the amplitudes from the tree level and loop
mechanisms removing the $\epsilon_A\cdot\epsilon$ factor.

On the other hand, giving the large width of the axial-vector
meson, particularly for the $a_1$, 
it is appropriate to fold the expression of the amplitude with the
mass distribution of the axial-vector resonance.
 Hence the final amplitude is
obtained from the expression

\be
\Gamma_{A\to P\gamma}= 
 {\cal N}^{-1}
\int_{(M_A-2\Gamma_A)^2}^{(M_A+2\Gamma_A)^2}
(-)\frac{ds_A}{\pi}  \,
Im \left\{\frac{1}{s_A-M_A^2+iM_A\Gamma_A}\right\}\,
 \Gamma(\sqrt{s_A})\,
\Theta(\sqrt{s_A}-\sqrt{s_A^{th}})
\label{eq:convolution}
\ee
\noindent
where $\Theta$ is the step function, $M_A$ and
$\Gamma_A$ are the nominal mass and total axial-vector
 meson width from the PDG \cite{pdg06} and $s_A^{th}$ is
the threshold for the dominant $A$ decay channels.
The errors quoted in the PDG for these magnitudes are taken into
account in the error analysis. In Eq.~(\ref{eq:convolution}), 
${\cal N}$ is a normalization factor in the convolution integral
obtained from the same integral as in 
Eq.~(\ref{eq:convolution}) setting $\Gamma(\sqrt{s_A})=1$.

Once the formalism and the different vertices have been exposed, we
are in a situation to address the possible contribution from
mechanisms involving the $\pi$-$a_1$ mixing. The mixing of
axial-vector and pseudoscalar mesons (or vectors and scalars) is
possible through the longitudinal component of the spin-1 propagator
$P^\mu P^\nu/P^2$ 
\cite{Ebert:1985kz,Nowakowski:1993iu,Veltman:1992tm,Kaloshin:1996kz,castro}.
In the Appendix we show that the diagrams corresponding to the present problem,
involving this mixing, vanish in our formalism.

\section{Results}

\begin{table}[h]
\begin{center}
\begin{tabular}{|c|c|c||c|}
\hline
 & &$\Gamma_{b_1\to\pi\gamma}$ & $\Gamma_{a_1\to\pi\gamma}$ \\ \hline

tree level  & $\phi$   & $23$ & $-$      \\
	    & $\omega$ & $16$ & $-$      \\
	    & $\rho$   & $-$   & $650$  \\ \cline{2-4}
	    & total    & $76$ & $650$  \\ \hline
loops       & type-b $K^*K$	   	 & $30$      & $10$   \\
      & type-b $\rho\pi$  	 & $-$       & $96$  \\
      & type-b  total    		 & $30$      & $136$  \\ \cline{2-4}
      & type-c $K^*K$	         & $0.14$	     & $0.05$   \\
      & type-c $\rho\pi$         & $-$      & $0.7$   \\ 
      & type-c  total     		 & $0.14$	     & $1.02$   \\ \cline{2-4}
      &  $K^*K$	                 & $34$      & $8.7$   \\
      &  $\rho\pi$   		 & $-$	     & $102$  \\ \cline{2-4}

       & total  loops                  & $34$      & $137$  \\  \hline \hline
total this work  & &$210\pm 40$ & $460\pm 100$\\ \hline
experiment & &$230\pm 60$ \cite{Collick:1985yi} & $630\pm 246$ \cite{Zielinski:1984au}\\ 
\hline
\end{tabular}
\caption{Different contributions to the radiative decay widths. All the
units are KeV.}
\label{tab:results}
\end{center}
\end{table}

In table~\ref{tab:results} we show the different contributions to
the partial decay width coming from the different mechanisms
considered in the calculation. The theoretical error in the final
results have been obtained by doing a Monte-Carlo sampling of the
parameters of the model within their uncertainty and considering
the uncertainties in the couplings discussed above. Note, however,
that we have no freedom in the  theory once the relevant parameters
(actually a subtraction constant) are fixed. 
  To
these  errors we add in quadrature  the 10\% from the arguments
used above concerning the $1/{m_V}^2$ terms.

From the results one can see that the tree level contribution for
the $a_1$ accounts for most of the decay width. However, for the
$b_1$ the tree level by itself only accounts for about $1/3$
of the experimental result, despite having two diagrams, $\phi$ and
$\omega$, that contribute to the tree level process. The smallness of
the tree level contribution comes from the $-\sqrt{2}/3$ and $1/3$
factor of the $\phi$ and $\omega$ coupling to the photon in comparison to the
 factor $1$ for the $\rho$ case  present in the $a_1$ decay and also
to the fact that the $a_1\rho\pi$
coupling obtained with the chiral unitary model \cite{Roca:2005nm}
is larger than the   $b_1\phi\pi$ and $b_1\omega\pi$.

Note also the constructive interference between the $\phi$ and
$\omega$ diagrams  despite the coefficient of the $V\gamma$
coupling having a different sign. This is so because  the
couplings of the $b_1$ to $\rho \pi$ and $\omega \pi$  have also
relative different sign. These relative signs are also
 a genuine non-trivial 
prediction of the  
chiral unitary model of Ref.~\cite{Roca:2005nm}.

Regarding the loop contribution, the total loop results
 for the $b_1$ and $a_1$ decays
have a comparable absolute value. In the $b_1$ case
 it 
increases the decay rate to account very well for the observed experimental
result, after interfering constructively with the
tree level mechanism. Note that  the most important contribution from the
loops comes from the type-b mechanisms 
(see Fig.~\ref{fig:diagrams}$b$). Particularly, for 
the $a_1$ case the dominant contribution to the loops come from the 
 $\rho\pi$ loops.

It is worth stressing the important role of the interferences among
the different terms to give the final result. The interferences
depend essentially on the sign of the couplings and the imaginary
part of the loop functions. The values and  relative signs of the
$AVP$ couplings are non-trivial predictions of the chiral unitary
model of Ref.~\cite{Roca:2005nm} and hence, the agreement of our
calculated radiative decay widths with experiment gives support
to the model  of Ref.~\cite{Roca:2005nm} and, hence, the
dynamical nature of these axial-vector resonances.

\section{Conclusions}

We have studied here the radiative decays  $a_1^+ \to \pi^+\gamma$
and  $b_1^+ \to \pi^+\gamma$ which had proved problematic before
in several approaches. The novelty which allowed us to obtain a
simultaneous good description of both decays  was the consideration
of the  $a_1$ and $b_1$ axial vectors as dynamically generated
resonances within the context of unitarized chiral perturbation
theory. Because of that we found important loop contributions that
were essential in reproducing the experimental values.
Technically, it is particularly rewarding to see that, by invoking
gauge invariance, the calculation is simplified and the relevant
loops are shown to be convergent despite their large superficial
degree of divergence.
  The nature
of the resonances as quasibound states of meson and vector-meson
states reverted into a loop contribution which provided a
substantial contribution to the radiative amplitudes, particularly
to the one of the $b_1$ radiative decay. 

One might think that loop
contributions could have been considered without resorting to the
concept of a dynamically generated resonance, by simply taking the
couplings of the resonance to their decay channels.  However, for
the important case of the $b_1$ we found that the contribution of
the
$\omega \pi$ loop vanished and the relevant contribution was coming
from $\bar{K} K^*$ and  $\bar{K^*} K$  which is a closed decay
channel (up to the  width of the states) and for which there is no
valuable experimental information. The chiral unitary approach
provides directly such couplings 
with definite signs
since these states are a part of
the building blocks of the resonances in the coupled channel
approach. 
Similarly, with the use of a phenomenological Lagrangian,
like the one of Ref.~\cite{palomar}, one could get such couplings,
but these are based on $SU(3)$ symmetry which 
is actually broken when
one generates dynamically resonances with a nonperturbative
approach like the one in Ref.~\cite{Roca:2005nm}. One example of
relevance to the present case is that, with the phenomenological
Lagrangian, the $b_1\to\phi\pi$ coupling is forbidden while in our
case, the nonperturbative treatment of the problem, involving many
iterative loops, generates a finite coupling that is dominant in
the tree level contribution of $b_1\to\pi\gamma$ 
(see Table~\ref{tab:results}).

The fact that we obtain a good description of the two
radiative decay rates for the first time provides support for the idea of
the axial vector mesons as dynamically generated states within
chiral dynamics. Other tests could follow as we get increased and
more accurate information on the axial vector mesons, and the
findings of the present work should serve to stimulate efforts in
this direction.

\section{Appendix: Mechanisms related to the mixing of axial-vector and
pseudoscalar mesons} 

\begin{figure}[h]
 \begin{center}
\includegraphics[width=0.8\textwidth]{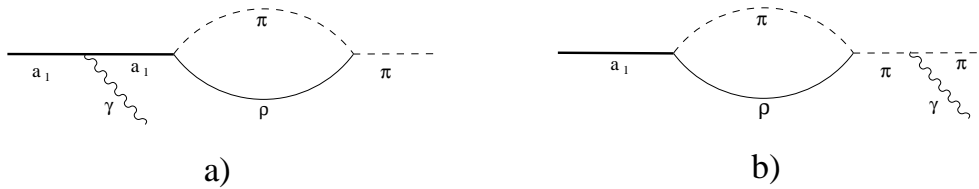}
\caption{Diagrams involving the $a_1$-$\pi$ mixing.} 
\label{fig:app1}
 \end{center}
\end{figure}

In addition to the terms discussed so far in this paper, we could have terms
involving the mixing of the axial-vector and pseudoscalar mesons
\cite{Ebert:1985kz,Nowakowski:1993iu,Veltman:1992tm,Kaloshin:1996kz,castro},
 through the longitudinal component of the axial-vector
resonance. In our case this occurs with $a_1$-$\pi$ mixing, allowed by
$G$-parity. The possible mechanisms involving this mixing in our scheme are
given in Fig.~\ref{fig:app1}.
However, we shall demonstrate here that they vanish in our formalism.

A free spin-1 massive meson propagator can be written as 

\begin{equation}
  \frac{-g^{\mu\nu}+\frac{P^\mu P^\nu}{M^2}}{P^2-M^2}
 =\frac{-g^{\mu\nu}+\frac{P^\mu P^\nu}{P^2}}{P^2-M^2}
+\frac{P^\mu P^\nu}{M^2 P^2}
\end{equation}
\noindent where in the second term of the equality a separation has been done in
terms of a transverse part ($-g^{\mu\nu}+\frac{P^\mu P^\nu}{P^2}$)
and a longitudinal one ($P^\mu P^\nu$). Note that the pole of the particle
appears only in the transverse part.

\begin{figure}[h]
 \begin{center}
\includegraphics[width=0.6\textwidth]{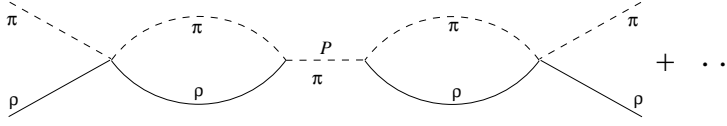}
\caption{Extra contributions to the $VP\to VP$ interaction involving the
$a_1$-$\pi$ mixing} 
\label{fig:app2}
 \end{center}
\end{figure}

In our formalism, the axial-vector resonance is dynamically generated from the
$VP$ interaction and is associated to the poles of the scattering matrix. In the
Appendix~B of Ref.~\cite{Roca:2005nm} we made explicitly the separation into
transverse and longitudinal part, with the result that the poles appeared only in
the transverse part of the amplitude. There we found for $T_{VP\to V'P'}$

\begin{equation}
T=\epsilon_\mu\epsilon'_\nu\left[\frac{Vb}{1-b}
\left(g^{\mu\nu}-
\frac{P^\mu P^\nu}{P^2}\right)
+\frac{Vc}{1-c}
\frac{P^\mu P^\nu}{P^2}\right]
\label{eq:TVPVPapp}
\end{equation}
\noindent 
where $P$ is the total momentum of the $VP$ system and $\epsilon$, $\epsilon'$,
the polarizations of the two vector mesons.
In 
Eq.~(\ref{eq:TVPVPapp}), $c$ is very small compared 
to $b$ and of opposite sign, such that there are no poles in
the longitudinal part. If we consider also $a_1$-$\pi$ mixing, we would have to
add terms like in Fig.~\ref{fig:app2} to our $VP$ amplitude.

The loop function appearing in Fig.~\ref{fig:app2} 
has the structure $J(P^2)P^\mu$. The sum of
terms in Fig.~\ref{fig:app2} renormalizes the longitudinal part of Eq.~(\ref{eq:TVPVPapp})
which is changed to 
\begin{equation}
\frac{Vc}{1-c - \frac{\beta J^2(P^2)}{P^2-m^2}}\frac{P^\mu P^\nu}{P^2}
=\frac{Vc(P^2-m^2)}{(1-c)(P^2-m^2)- \beta J^2(P^2)}\frac{P^\mu P^\nu}{P^2}
\end{equation}
where $m$ is the pion mass. The amplitude has the unphysical feature of
providing a pole related to the pion pole (close to $m^2$ if $\beta$ is small).
The way to remove this unphysical behaviour of the longitudinal part is to
demand that $J(P^2=m^2)=0$, which also appears in other formalism 
\cite{Kaloshin:1996kz}. In other works \cite{castro} it is shown explicitly that
the renormalized full vector meson propagator contains only one pole which does
not show up in the longitudinal part.

\begin{figure}[h]
 \begin{center}
\includegraphics[width=0.8\textwidth]{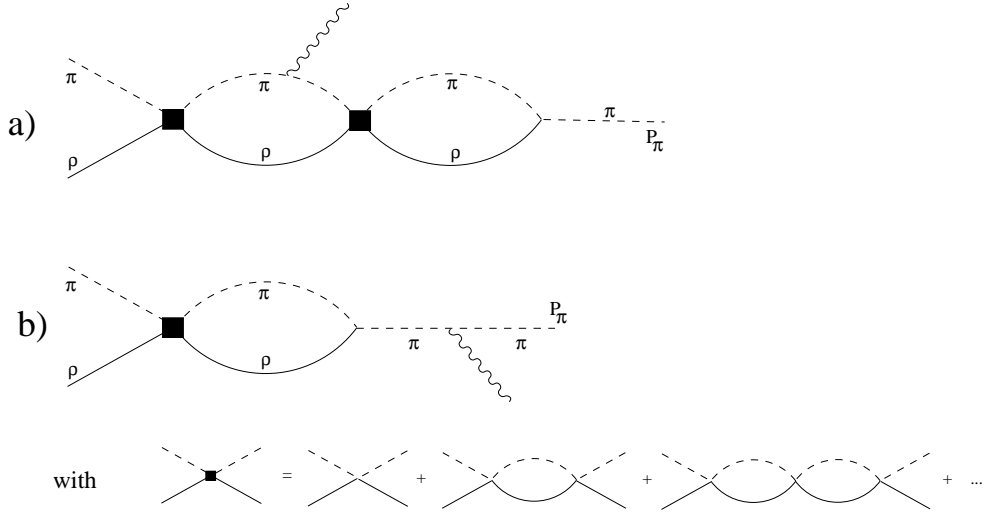}
\caption{Mechanisms of Fig.~\ref{fig:app1} in the dynamical formalism} 
\label{fig:app3}
 \end{center}
\end{figure}

The contribution of the 
mechanisms of Fig.~\ref{fig:app1} in our formalism would have to be considered 
through the
$a_1$ pole  of the amplitudes of Fig.~\ref{fig:app3}

Diagram Fig.~\ref{fig:app3}a), with the photon emitted either from the pseudoscalar or the
vector in the loop, is proportional to $J(P_\pi^2=m^2)$ and hence vanishes. 

Diagram \ref{fig:app3}b) is more subtle. The amplitude is proportional to
\begin{eqnarray}
&&\epsilon_\mu(\rho)\left[\frac{Vb}{1-b}\left(g^{\mu\nu}-
\frac{P^\mu P^\nu}{P^2}\right)
+\frac{Vc}{1-c- \frac{\beta J^2(P^2)}{P^2-m^2}}
\frac{P^\mu P^\nu}{P^2}\right]J(P^2)P^\nu \nonumber \\
&=&\epsilon_\mu(\rho) P^\mu J(P^2)\frac{Vc}{1-c- \frac{\beta J^2(P^2)}{P^2-m^2}}
\label{eq:appxxx}
\end{eqnarray}
\noindent which has filtered the longitudinal part of the amplitude.
 Furthermore, the procedure we have followed to evaluate the coupling of $a_1$
to $\pi\gamma$ is equivalent to calculating the residue of the
$\pi\rho\to\pi\gamma$ amplitude and dividing by the $a_1\to\pi\rho$ coupling
\cite{Hyodo:2006uw}. The residue of Eq.~(\ref{eq:appxxx})
 at the resonance pole ($b=1$) is
zero and hence the mechanism of Fig.~\ref{fig:app3}b)
 also vanishes at the $a_1$ pole.

\section*{Acknowledgments}  

We would like to express our thanks to Hideko Nagahiro who checked
our calculations and help us to improve them.
This work is partly supported by DGICYT contract
number FIS2006-03438, the Generalitat Valenciana and 
the JSPS-CSIC collaboration agreement no. 2005JP0002.
This research is 
 part of the EU Integrated
Infrastructure Initiative  Hadron Physics Project under  contract number
RII3-CT-2004-506078.

\end{document}